# Representative Scenarios to Capture Renewable Generation Stochasticity and Cross-Correlations

Dhaval Dalal, *Senior Member, IEEE*
Anamitra Pal, *Senior Member, IEEE*
School of Electrical, Computer and Energy Engineering
Arizona State University, Tempe, AZ, USA
ddalal2@asu.edu, Anamitra.Pal@asu.edu

Philip Augustin, *Member, IEEE*
Transmission Analysis,
Salt River Project (SRP)
Phoenix, AZ, USA
Philip.Augustin@srpnet.com

***Abstract*—Generating representative scenarios for power system planning in which the stochasticity of renewable generation and cross-correlations between renewables and load are fully captured, is a challenging problem. Traditional methods for scenario generation often fail to generate diverse scenarios that include both seasonal (frequently occurring) and atypical (extreme) days required for planning purposes. This paper presents a methodical approach to generate representative scenarios. It also proposes new metrics that are more relevant for evaluating the generated scenarios from an applications perspective. When applied to historical data from a power utility, the proposed approach resulted in scenarios that included a good mix of seasonal and atypical days. The results also demonstrated pertinence of the proposed cluster validation metrics. Finally, the paper presents a trade-off for determining optimal number of scenarios for a given application.**

## I. Introduction

As power systems start incorporating more renewable resources into their fleet, the planning and operations activities must be altered as well. Traditional planning concentrated in understanding load patterns (usually based on historical data). However, renewable resources bring their own variability and uncertainty in the mix, requiring these attributes to be fully considered in long- and medium-term planning activities such as load growth, new investments, storage planning, reliability, and outage coordination. Scenario-based solutions are often utilized in such activities to keep the scale of the problems manageable while still giving accurate results. With renewables, the scenario generation task becomes more complex due to (a) addition of more variables, and (b) need to capture cross-correlations between the variables [1], [2].

Scenario generation can be based on probabilistic models [3], [4] or based on mining of historical data [5], [6]. However, generation of a multivariate probabilistic model representing all correlations is extremely challenging [6]. Clustering techniques are commonly used to generate scenarios based on historical data. Traditional clustering algorithms such as K-Means and K-Medoids rely on Euclidean distance between the time series to create clusters of similar days where the mean or centroid of each cluster denotes the representative day, and the number of days in each cluster represents the likelihood of that scenario. The clustering methods cover a wide range of end applications, including capacity expansion, unit commitment, transmission line capacity planning, and storage planning [3]-[7]. For most of these applications, the value of clustering is demonstrated by comparing the final outcome realized using the original data and the representative data. However, this does not make a compelling case for choosing a scenario generation approach for a given application. There are also concerns regarding the qualitative or metric-based appropriateness of the approach based on the clustering results. Lastly, note that as clustering is unsupervised classification, cluster validation is often "the most difficult and frustrating part of cluster analysis" [8].

In this paper, we try to address this challenging problem by presenting a more methodical approach that identifies the most suitable clustering technique based on the application requirements. Particularly for the case of renewable generation and loads, we focus on generating scenarios that have a good mix of typical days (seasonal) and atypical days (where one or more of the variables has an abnormal pattern). For representative scenario generation for planning problems, we identify two important requirements: (a) getting the right measure of similarity, and (b) adequately capturing the diversity. This paper applies dynamic time warping (DTW) to multivariate time series to get a more relevant similarity measure between different days for enabling better clustering. Subsequently, the impact of clustering approaches on the desired goal of representative day diversity is explored with pertinent results and analyses, which include, for the first time in the authors' knowledge, histograms of monthly population density in each cluster. Finally, a new pair of metrics is proposed for cluster validation so that the right clustering technique and correct number of clusters can be chosen at the clustering output level. This reduces the need to do elaborate planning exercises with different sets of scenarios. In summary, the major contributions of this paper are as follows:

- Linking the similarity measure and clustering technique choices to the application requirements

This work was supported in part by Salt River Project (SRP) under grant 96-180C 2021-2022 EE-04.

- Establishing a methodical approach to create and analyze representative load and renewable generation scenarios using monthly population density
- Developing novel metrics that allow accurate cluster validation
- Methodology for selecting optimal number of clusters for a given application

## II. Dynamic Time warping (DTW) and Agglomerative Hierarchical Clustering (AHC)

Main attributes of DTW and AHC, and preliminary results obtained using a combination of these two techniques, are covered in this section.

### *A. Dynamic Time Warping (DTW)*

DTW facilitates better determination of similarity measures between time series data compared to Euclidean distance [9]. It computes the best alignment between two time series by identifying the path with minimum time-normalized distance between them. This is given by (1).

$$P_0 = arg\min_P \left[\frac{\sum_{s=1}^{k} d(p_s)}{k}\right] \quad (1)$$

where $d(p_s)$ is the distance between time series points $i_s$ and $j_s$, $k$ is the length of the warping path and $P$ is the warping function. For univariate time series data (e.g., hourly solar profile per day), the time series $i_s$, $j_s \in \mathbb{R}^{24\times1}$ and the DTW operation identifies the smallest distance by permuting through the different paths from hour 1 to hour 24. When considering multivariate time series data, each day is represented by a matrix $d_i \in \mathbb{R}^{24\times m}$ (where $m$ is the number of variables) and the DTW operation is performed for each variable. The importance of DTW for multivariate renewable applications is that it is able to better capture the cross-correlations between variables (such as load and solar generation), and it identifies similarities between patterns even if they are time-displaced. For $n$ days, the DTW operation creates a symmetrical $\mathbb{R}^{n\times n}$ matrix that has the DTW distances between each pair of days.

### *B. Agglomerative Hierarchical Clustering (AHC)*

To create representative days from the outputs of DTW, clustering algorithms can be employed. However, traditional clustering techniques such as K-Means do not work well with non-globular clusters and have difficulty handling outliers [10]. Therefore, AHC [11] is used here for creating representative days. Since clustering is an unsupervised learning algorithm, there is no direct measure of its effectiveness. Hence, the attributes used for AHC, such as linkage choices, must be chosen keeping in mind the objectives of the end application. Another aspect of clustering is determination of the number of representative scenarios that effectively represent the complete dataset. Higher number of scenarios may be more accurate, but less useful due to heavier computational burden. This issue is addressed in detail in Section III.

For planning applications, both typical and atypical days must be adequately captured in the representative scenarios, while not making the number of scenarios too large. In this respect, two AHC linkage choices, namely, complete and average, are investigated. Complete linkage tries to incorporate outliers into more normal clusters, while average linkage can create separate clusters consisting of distinct outliers [12]. When a fixed number of clusters is used, this results in different densities of normal days.

### *C. Preliminary Analysis*

DTW and AHC was applied to 2-years' worth of historical data for solar generation and load obtained from the Salt River Project (SRP), a power utility in the U.S. Southwest. Preliminary results obtained from traditional clustering approaches (K-Means) based on Euclidean distance (Figure 1) were compared with those obtained using the proposed approach (Figure 2, Figure 3). All three figures have 14 representative scenarios with normalized hourly load data (top two rows), normalized hourly solar data (middle two rows), and histograms of monthly population density captured in each cluster (bottom two rows). The first two rows also identify the cluster number and the number of days in each cluster. Colored plots are the raw data, while the black lines indicate the cluster representations in each of the first four rows. The following observations are made from these three figures:

- Although K-Means clusters represented some of the average behavior better, there were no seasonal/typical clusters identified using it. This is because K-Means clustering distributed the days fairly evenly across all the clusters. Similarly, no clusters showed outlier information in the results obtained using K-Means (the smallest cluster had 16 days). As such, this type of clustering yields very little useful information for planning purposes where a balance between similarity and diversity is needed. Another aspect of K-Means clustering is that it is not repeatable – every run produces a different set of clusters.
- With complete linkage AHC, some seasonal clusters emerge (e.g., cluster 1 for summer, cluster 7 for winter), but the density of the seasonal clusters is still low. Complete linkage also displays some outlier clusters (e.g. clusters 9 and 11). Based on the available data, these tend to be dominated by stochasticity of solar generation.
- Average linkage AHC has a more complete seasonal representation (cluster 2 for spring, cluster 6 for winter, cluster 9 for summer). Moreover, 8 of the 12 months have more than 2/3 of their days in a single cluster, indicating more coherent clustering. Average linkage also does a better job identifying outliers and putting them in smaller clusters (often with single member) – *this can be useful for planning operations involving reliability assessments*. Lastly, average linkage is able to distinguish between load and solar variations. Clusters 12 and 13 that have similar solar patterns, but the load is much higher in cluster 12 in comparison to cluster 13.

The above-mentioned observations, while made by analyzing a 2-year dataset and using 14 clusters, were found to be qualitatively consistent across other combinations of datasets and representative days.

## III. Metrics for Cluster validation

Since clustering is an unsupervised learning method, its effectiveness cannot be quantified easily. One option is to apply the clustering results to the final objective (typically an optimization algorithm) and compare the results with those obtained by using unclustered data. However, that usually involves heavy computational burden. In addition, with multiplicity of clustering algorithms and a range of cluster numbers, there is a need to add metric-based selection at the clustering level. The previous section showed a methodology for evaluating the clustering outcomes based on the analysis of the representative plots. However, that methodology is also not scalable to different applications.

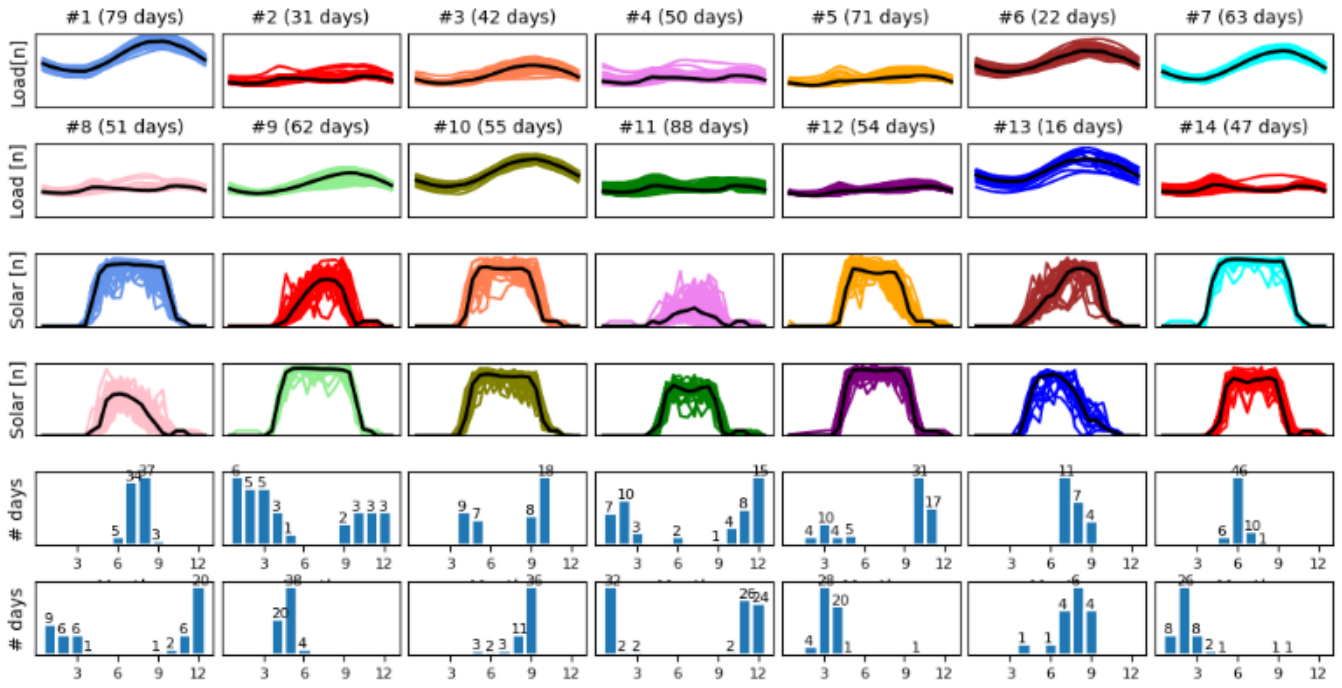

*Figure 1. Representative days with K-Means clustering*

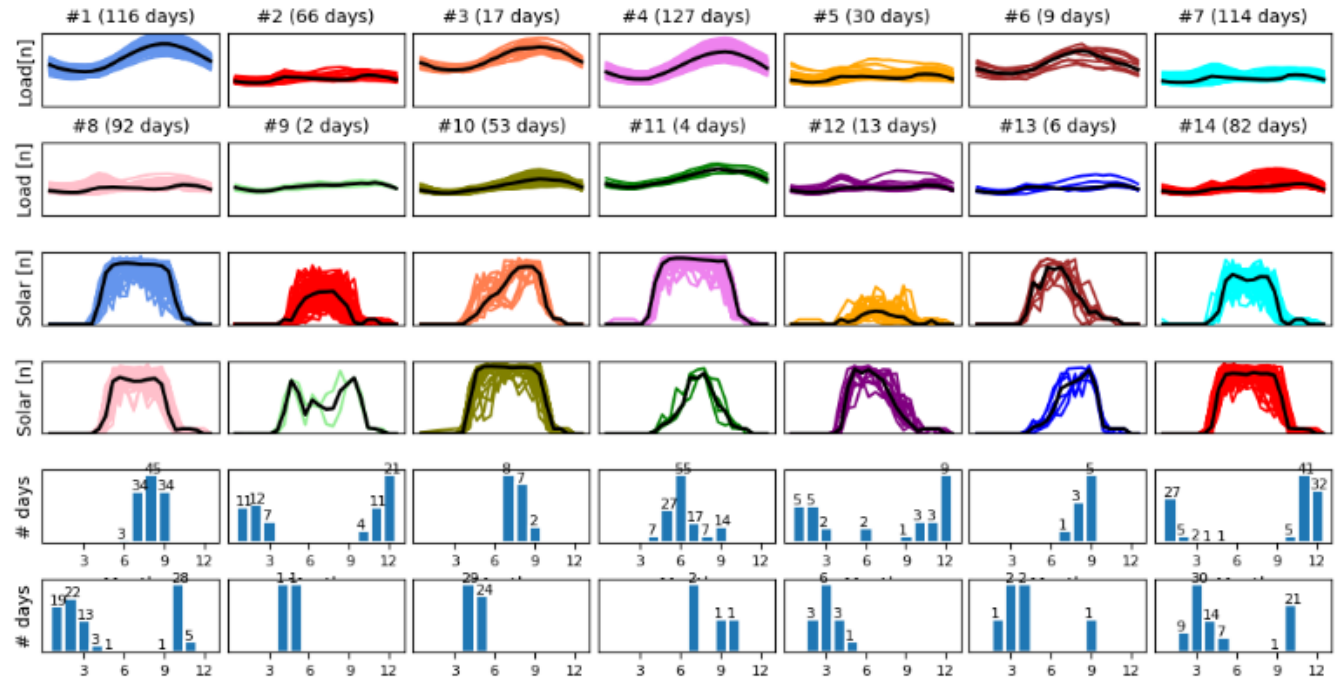

*Figure 2. Representative days with DTW+AHC (Complete linkage)*

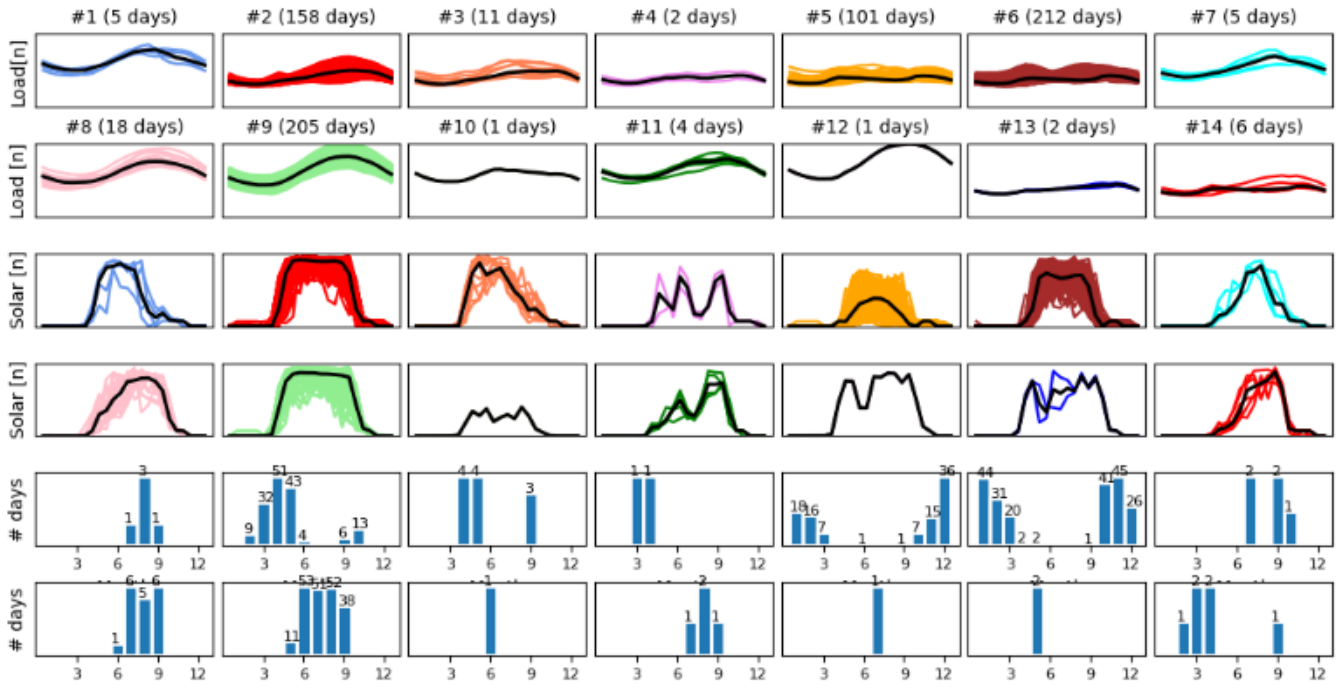

*Figure 3. Representative days with DTW+AHC (Average linkage)*

### *A. Basic Properties of Clustered Data*

When data are clustered into different representations, there are two opposing criteria that must be satisfied to achieve optimal clustering. One is the *intra-cluster cohesion* and the other is the *inter-cluster separation*. With inherently clean data (with uniform density), it is easy to satisfy both criteria effectively. However, with multivariate, stochastic data, there is a balancing act wherein improving one measure (e.g. intra-cluster cohesion) leads to deterioration of the other (namely, inter-cluster separation). One available variable to achieve the right balance is the number of clusters. Intuitively, it is easy to see that low number of clusters can ensure better inter-cluster separation while sacrificing intra-cluster cohesion and the opposite is true for high number of clusters. However, the results can also differ based on the clustering algorithm employed. In the results shown in the previous section, it was evident that K-Means clustering (Figure 1) rewarded intra-cluster cohesion more than inter-cluster separation.

### *B. Available Metrics and Their Limitations*

Many metrics for evaluating clustering effectiveness are already available; examples include, Calinski-Harabasz (CH) index, Davies-Bouldin (DB) index, and Silhouette score [13]. However, they are defined around Euclidean distances. As explained in the previous section, the similarity score based on DTW distances is more relevant for renewable planning applications. Hence, the clustering effectiveness should also be evaluated based on the DTW distances. In addition, all these metrics provide a single numerical value which makes built-in assumptions (that may not be universally valid) about the trade-offs between the two criteria mentioned in Section III.A. For example, these metrics may reward the inter-cluster separation over intra-cluster cohesion and hence, suggest that a lower number of clusters (often two clusters) is the best option. However, considering the variability in renewable energy resources, it is clear that the optimum number of clusters for satisfying all the requirements (e.g., seasonality, outlier days, cloudy days) can be much higher [14]. This leads us to propose new metrics for cluster validation.

### *C. Proposed Metrics*

To develop a clear, consistent methodology for cluster validation, two new metrics are proposed that address the issues identified with existing metrics, namely, reliance on Euclidean distances, and built-in weightage between the two opposing criteria mentioned in Section III.A. One metric each is proposed for intra-cluster cohesion and inter-cluster separation criteria. Let the distance of a day to its cluster centroid be given by (2):

$$MA_i = D_{DTW}(d_i, c_k) \mid i \in p_k \tag{2}$$

Then, distance of a day to its nearest cluster is given by (3):

$$MC_i = \min_k [D_{DTW}(d_i, c_k)] \mid i \notin p_k \tag{3}$$

where for a dataset of $N$ days and $K$ clusters,; $d_i$ is the multivariate time series data for a day $i \in [1, N]$, $c_k$ is the multivariate time series representation (centroid) of a cluster $k \in [1, K]$, $p_k$ is the membership list of days in cluster $k$, and $D_{DTW}(x, y)$ is the DTW distance between series $x$ and $y$. The difference between $MC_i$ and $MA_i$ represents a more accurate measure of inter-cluster separation because if a day $i$ is closer to the other cluster centroid than its own centroid, the difference would be negative. On the other hand, if there is sufficient separation, the difference would be positive and large. Hence, the first proposed metric, *separation score* (SS), for inter-cluster separation is given by (4):

$$SS = \frac{\sum_{i=1}^{N}(MC_i - MA_i)}{N} \tag{4}$$

Similarly, when the highest value of $MA_i$ is low within a cluster, it represents better intra-cluster cohesion. Hence, the second proposed metric, *cohesion score* (CS), is given by (5):

$$CS = \max_{k}[\max_{p_k}(MA_i) \mid i \in p_k] \quad (5)$$

In addition to incorporating DTW distance, each of the proposed metrics incorporates measurement of an individual day against the cluster centroids (its own as well as other clusters). The existing metrics either use day-day distances (Silhouette score) or a combination of day-cluster and cluster-cluster distances (CH and DB indices). This difference in approaches is important when considering that the end objective of the representative scenarios is to evaluate the output (cluster centroids) against the input (individual days). Consequently, the proposed metrics contain aggregated measures for each instance of the original dataset against the eventual representation, which results in a more relevant metric for scenario generation. The day-day distances and cluster-cluster distances are less meaningful in this context.

## IV. RESULTS

The proposed metrics were computed for a dataset consisting of 2 years of hourly solar generation and load data from the Salt River Project (SRP). Comparisons were made with the traditional metrics (CH index, DB index, and silhouette score) for measuring clustering effectiveness as well as across different clustering techniques and cluster numbers. These results are shown in Figure 4 (CH index), Figure 5 (DB index), Figure 6 (Silhouette score), Figure 7 (SS), and Figure 8 (CS) – the x-axis shows the number of clusters in each plot. The following observations are made from these figures:

- The three traditional metrics show that the K-Means algorithm is better than the AHC with DTW in many instances. However, there is no consistency in these metrics as the number of clusters is varied. As illustrated in Figure 1-Figure 3, the DTW methods provide more suitable representation for the given data and the traditional metrics fail to support that observation.
- SS shows that average linkage AHC performs better than all other techniques for any reasonable cluster size. It also captures the inability of complete linkage to separate out the outliers.
- The general trend for the traditional metrics and SS is to indicate that a smaller number of clusters is optimal. However, from an applications perspective it may be better to obtain a larger cluster size to provide seasonal as well as outlier days' information. Hence, the traditional metrics as well as SS alone are not always suitable.
- CS shows that both average and complete linkage AHC with DTW outperform traditional techniques, which is consistent with Figure 1-Figure 3.

The plots also show that while one metric (CS) improves as number of clusters increases, the other metric (SS) gets worse. This trade-off is inherently present when deciding on the optimal cluster size for any dataset. However, the optimum number of clusters should not be dictated by a single metric that predetermines the trade-off value between the two trends. Instead, the nature of the dataset and the application requirements must be considered to determine the optimal trade-off. As an example, Figure 8 shows that CS achieves its minimum value (within the cluster range of 2-20), at 14 clusters for both complete and average linkages and stays flat after that. Hence, there is no cohesion improvement between 14 and 20 clusters. On the other hand, SS has a flat region between 10 and 15 clusters. Thus, combining the two, it may be concluded that the choice of 14 or 15 clusters is optimum for this dataset. Further validation was carried out with other datasets which had an additional variable (namely, wind) and different time durations. The inferences were found to be consistent with the new datasets as well.

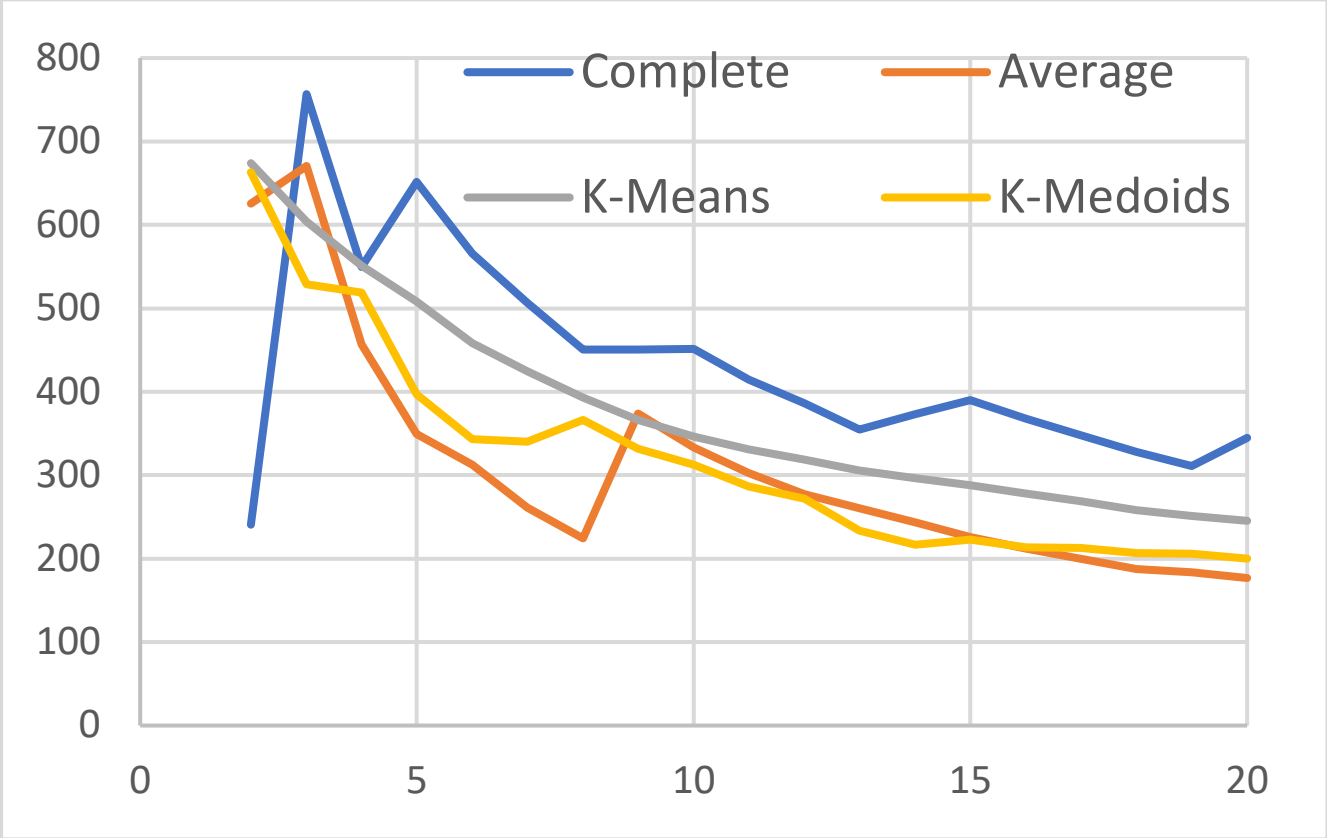

*Figure 4. Calinski-Harabasz Index (higher the better)*

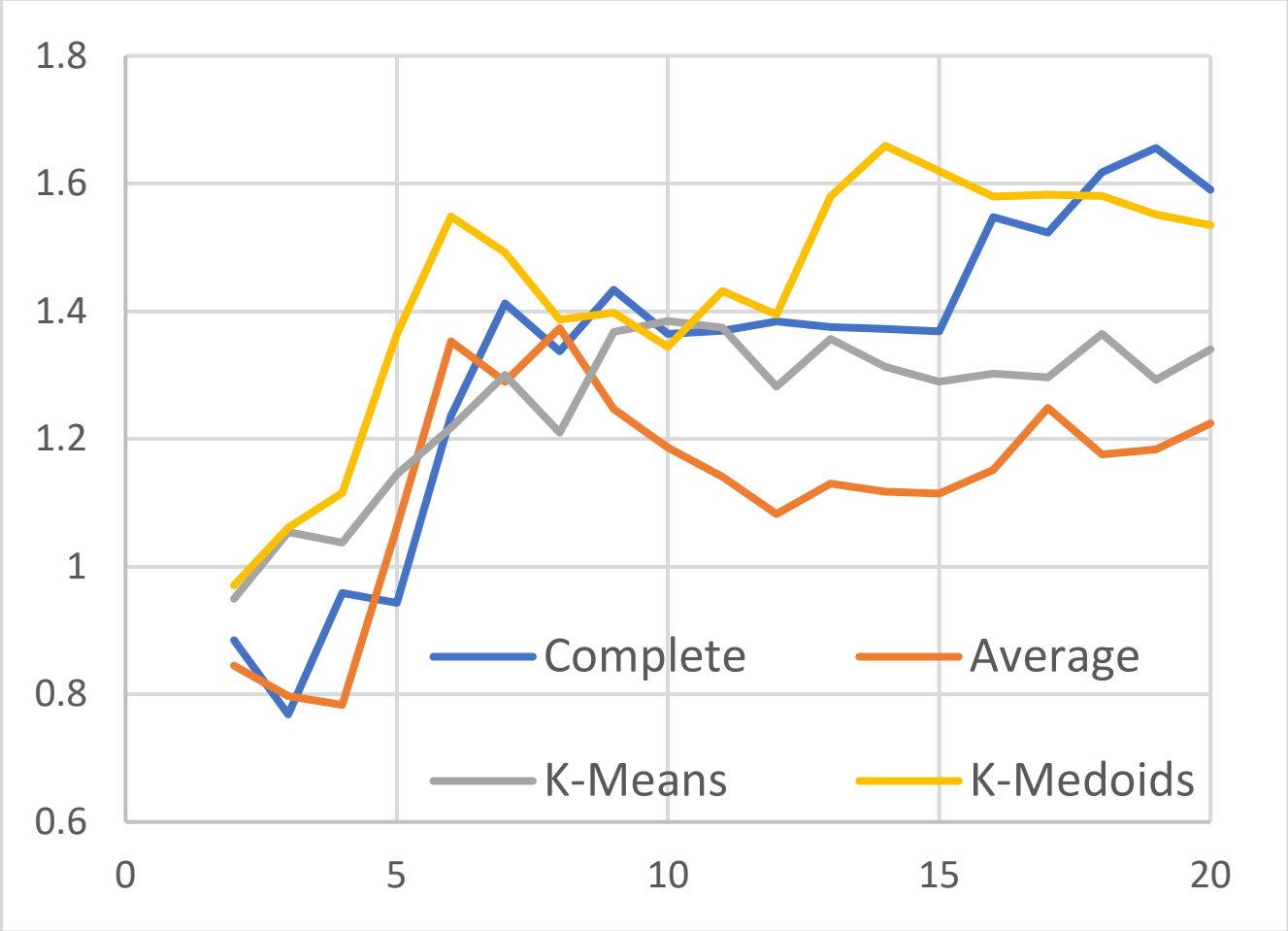

*Figure 5. Davies-Bouldin Index (lower the better)*

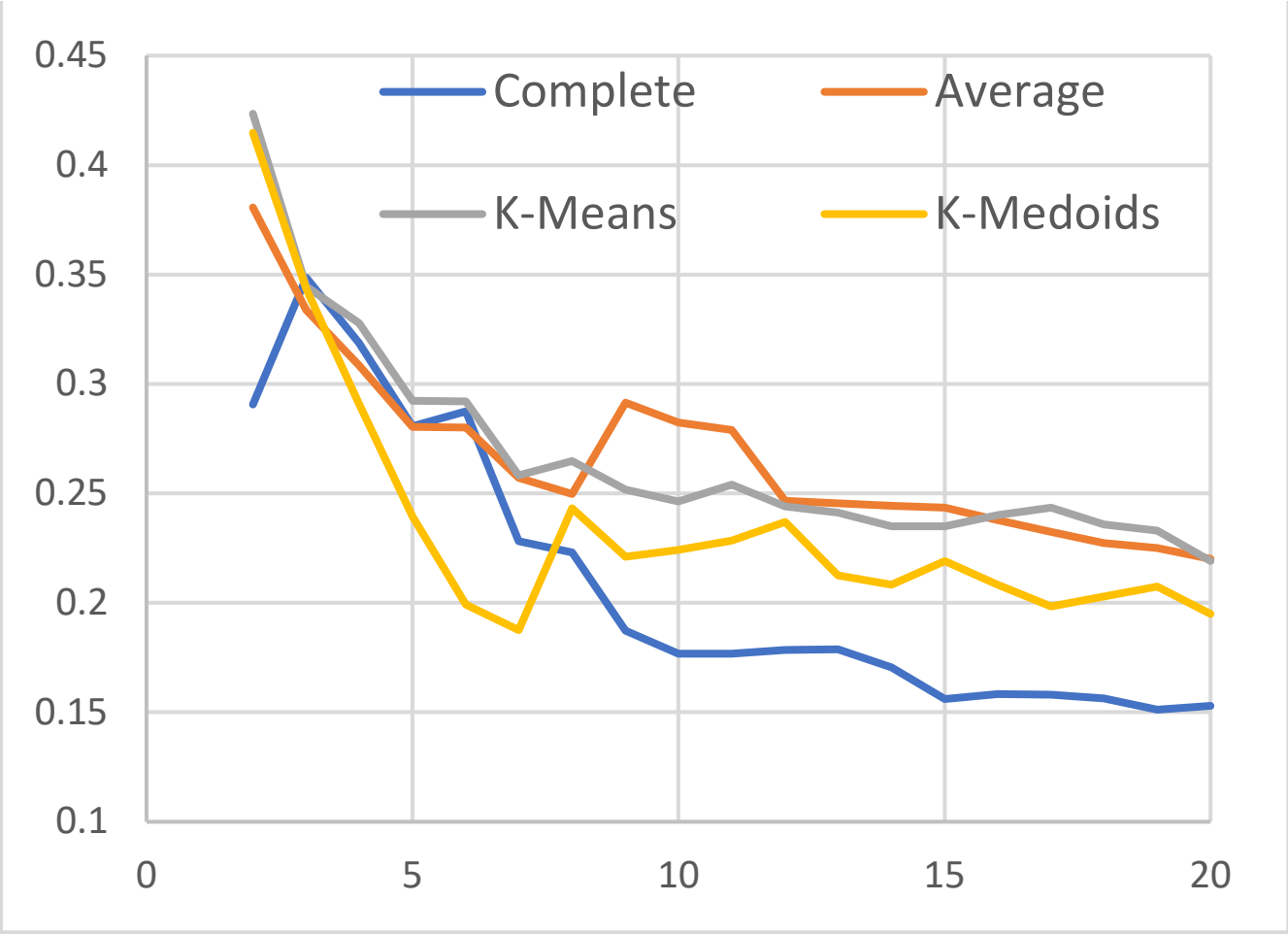

*Figure 6. Silhouette Score (higher the better)*

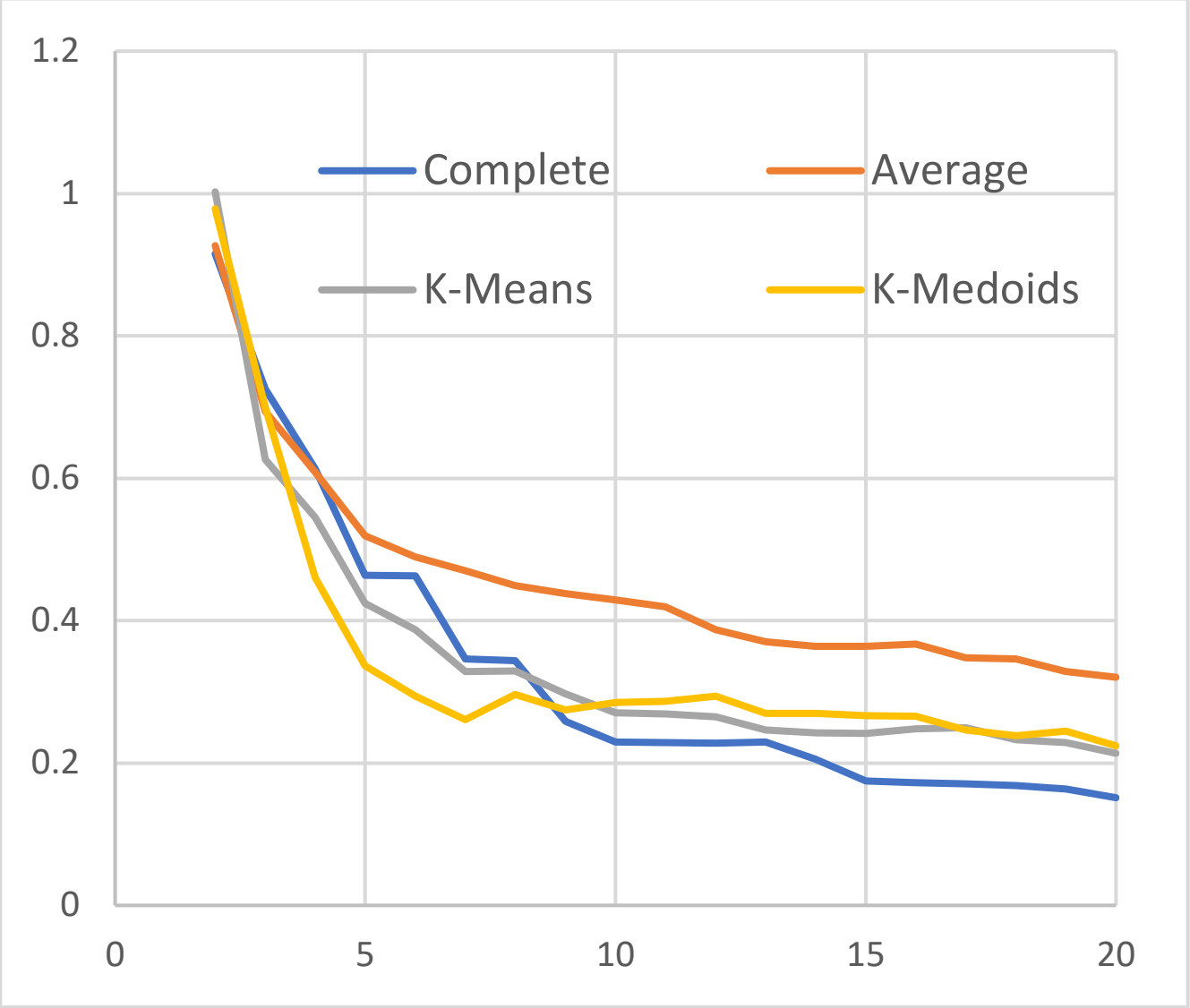

*Figure 7. Separation Score (higher the better)*

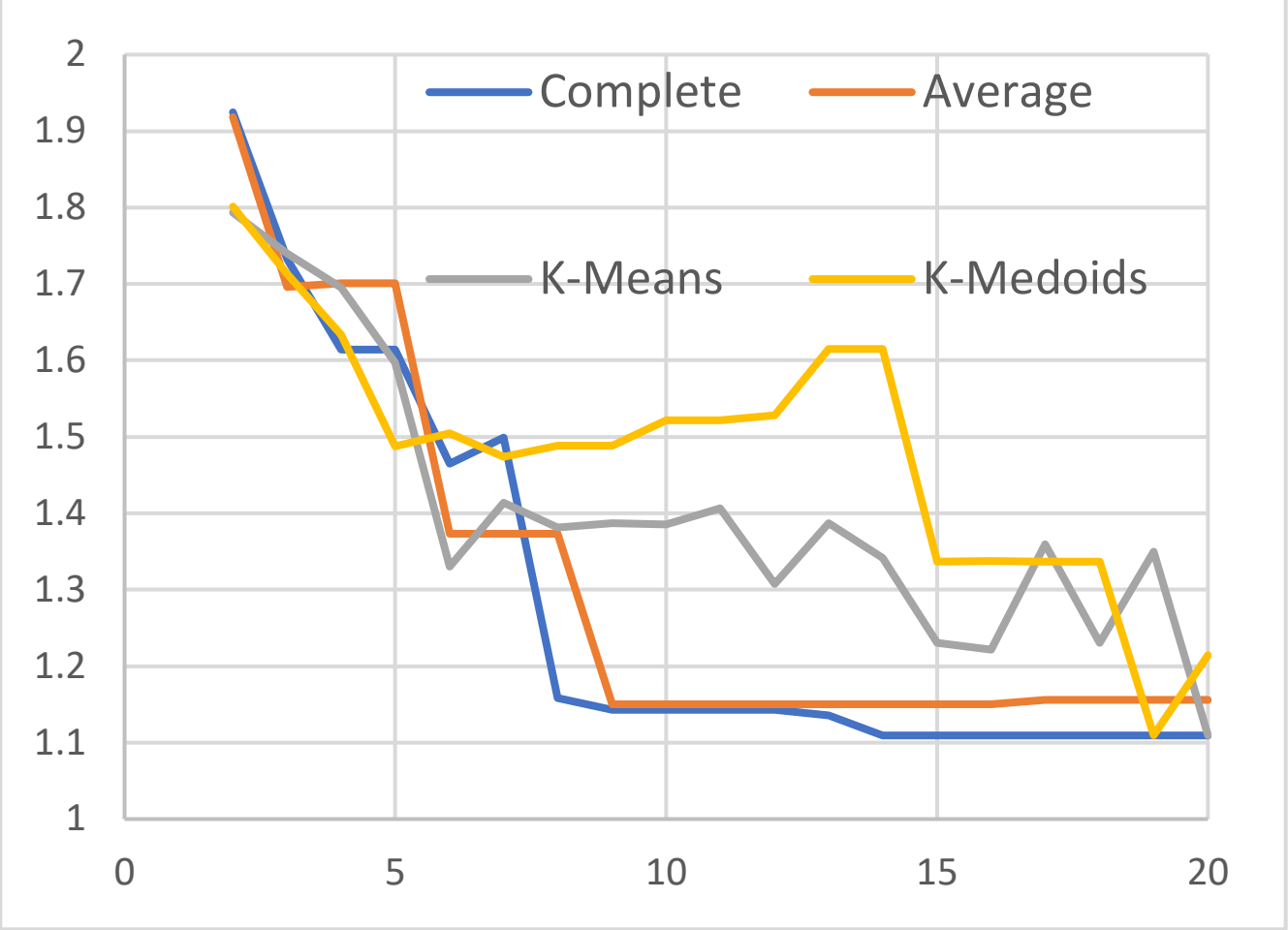

*Figure 8. Cohesion Score (lower the better)*

## V. CONCLUSIONS

This paper provided a framework for generating representative scenarios from stochastic data that contain variabilities and uncertainties associated with renewable generation. It demonstrated that the combined application of DTW and AHC resulted in valid representative scenarios. The validity of the scenarios was demonstrated based on a qualitative analysis as well as with new quantitative metrics proposed in this paper. The qualitative analysis used a visualization approach that combined diurnal patterns within the clusters with monthly population density histograms. This visualization helped conclude that the DTW+AHC with average linkage approach met the application requirement of generating a mix of seasonal typical as well as outlier days' representations.

The paper also showed how and why the conventional metrics for cluster validation fail to identify the more suitable clustering techniques as well as optimal number of clusters for a given application. It identified a more appropriate pair of statistical metrics to apply to clustering results for cluster validation and showed results to support this conclusion. Finally, the paper provided a trade-off-based evaluation methodology that can lead to the right number of clusters for a given application and for a given dataset.

The proposed generalized framework can be used to create scenarios for different applications and for a variety of datasets. It is also amenable to further fine tuning as required. The use of this framework for generating representative scenarios for solving capacity/storage planning problems will be explored in the future.

## REFERENCES

[1] M. Padhee, and A. Pal, "Effect of solar PV penetration on residential energy consumption pattern," in *Proc. IEEE North American Power Symposium (NAPS)*, Fargo, ND, pp. 1-6, 9-11 Sep. 2018.

[2] M. Padhee, A. Pal, and K. A. Vance, "Analyzing effects of seasonal variations in wind generation and load on voltage profiles," in *Proc. IEEE North American Power Symposium (NAPS)*, Morgantown, WV, pp. 1-6, 17-19 Sep. 2017.

[3] S. Camal, F. Teng, A. Michiorri, G. Kariniotakis, and L. Badesa, "Scenario generation of aggregated wind, photovoltaics and small hydro production for power systems applications," *Applied Energy*, vol. 242, pp. 1396–1406, May 2019.

[4] D. Ji and H. Wang, "A scenario probability based method to solve unit commitment of large scale energy storage system and thermal generation in high wind power penetration level system," in *Proc. IEEE PES Asia-Pacific Power & Energy Engineering Conf. (APPEEC)*, pp. 84–88, 2016.

[5] Y. Liu, R. Sioshansi, and A. J. Conejo, "Hierarchical clustering to find representative operating periods for capacity-expansion modeling," *IEEE Trans. Power Syst.* , vol. 33, no. 3, pp. 3029-3039, May 2018.

[6] J. Yang, S. Zhang, Y. Xiang, J. Liu, J. Liu, X. Han, and F. Teng, "LSTM auto-encoder based representative scenario generation method for hybrid hydro-PV power system," *IET Generation, Transmission & Distribution,* vol. 14, no. 24, pp. 5935-5943, Dec. 2020.

[7] B. Jiao, W. Chengshan, and L. Guo, "Scenario generation for energy storage system design in stand-alone microgrids," *Energy Procedia*, vol. 61, pp. 824–828, Nov. 2014.

[8] A. K. Jain, and R. C. Dubes, *Algorithms for Clustering Data*, Englewood Cliffs, NJ: Prentice Hall, 1988.

[9] C. A. Ratanamahatana, and E. Keogh, "Making time-series classification more accurate using learned constraints," in *Proceedings of the SIAM International Conference on Data Mining*, Philadelphia, PA: Society for Industrial and Applied Mathematics, 2004, pp. 11–22.

[10] M. K. Tripathi, A. Nath, T. P. Singh, A. S. Ethayathulla, and P. Kaur, "Evolving scenario of big data and artificial intelligence (AI) in drug discovery," *Molecular Diversity*, vol. 25, no. 3, pp. 1439–1460, Jun. 2021.

[11] C. Manning, P. Ragahavan, and H. Schutze, (2009), *Introduction to Information Retrieval* [Online], "Hierarchical agglomerative clustering," Available: *https://nlp.stanford.edu/IR-book/html/htmledition/hierarchical-agglomerative-clustering-1.html*. [Accessed 11 3 2021]

[12] C. Manning, P. Ragahavan, and H. Schutze, (2009), *Introduction to Information Retrieval* [Online], "Single-link and complete-link clustering," Available: *https://nlp.stanford.edu/IR-book/html/htmledition/single-link-and-complete-link-clustering-1.html.* [Accessed 11 3 2021]

[13] E. Zuccarelli. (2021, Jan. 31), *Towards Data Science* [Online]. "Performance metrics in machine learning — part 3: clustering", Available: *https://towardsdatascience.com/performance-metrics-in-machine-learning-part-3-clustering-d69550662dc6.* [Accessed 11 5 2021]

[14] M. Padhee, A. Pal, C. Mishra, and K. A. Vance, "A fixed-flexible BESS allocation scheme for transmission networks considering uncertainties," *IEEE Trans. Sustainable Energy*, vol. 11, no. 3, pp. 1883-1897, Jul. 2020.